\documentclass[letterpaper,10pt]{article} 

\usepackage{opticameet3} 

\newcommand\authormark[1]{\textsuperscript{#1}}

\usepackage{amsmath,amssymb}
\usepackage{graphicx}
\usepackage[colorlinks=true,bookmarks=false,citecolor=blue,urlcolor=blue]{hyperref} 
\usepackage{float}
\usepackage[dvipsnames]{xcolor}
\begin{document}

\title{Storage Buffer of Polarization Quantum States Based on a Poled-Fiber Phase Modulator} 


\author{Daniel Spegel-Lexne\authormark{1,*}, João Manoel Barbosa Pereira\authormark{2}, Alvaro Alarcón\authormark{3}, Joakim Argillander \authormark{1}, Martin Clason\authormark{1}, Åsa Claesson\authormark{4}, Kenny Hey Tow\authormark{4}, Walter Margulis\authormark{5} and Guilherme B. Xavier\authormark{1} }

\address{\authormark{1} Institutionen för Systemteknik, Linköpings Universitet, 581 83, Linköping, Sweden \\
\authormark{2} Fiber Optics, RISE-Research Institutes of Sweden, Electrum 236, 16440, Kista, Sweden \\
\authormark{3} Departmento de Ingeniería Eléctrica y Electrónica, Facultad de Ingeniería, Universidad del Bío-Bío, 4051381 Avenida Collao, Concepción 1202, Chile \\
\authormark{4} Fiberlab, RISE-Research Institutes of Sweden, Fibervägen 2-6, 82450, Hudiksvall, Sweden \\
\authormark{5} Pontifical Catholic University of Rio de Janeiro, 22451-190, Rio de Janeiro, Brazil}
\email{\authormark{*}daniel.spegel-lexne@liu.se} 

\begin{abstract}
Dynamic storage of qubits is crucial for quantum communication networks.
Here we present an adjustable buffer capable of storing photonic polarization quantum states in a fiber loop controllable by a poled fiber modulator.

\end{abstract}

\section{Introduction}
One of the great challenges for future quantum communication networks is the need for quantum memories. One type of quantum memory that shows promising results is a loop based one, which is basically a variable optical delay line \cite{Evans2023}. However, for simpler compatibility into commercial quantum communication systems, an all-fiber configuration would be the best solution. In our previous work, we implemented a Sagnac loop configuration with a poled fiber phase modulator to store classical light pulses \cite{Pereira24}, and in this work we develop a new poled modulator with 0.4 dB loss and demonstrate its ability to store and retrieve quantum information encoded on polarization photonic quantum states. 
\section{Experimental Setup and Results}

\begin{figure}[H]
    \centering
    \includegraphics[trim = 0 5 0 9,clip,width=0.93\linewidth]{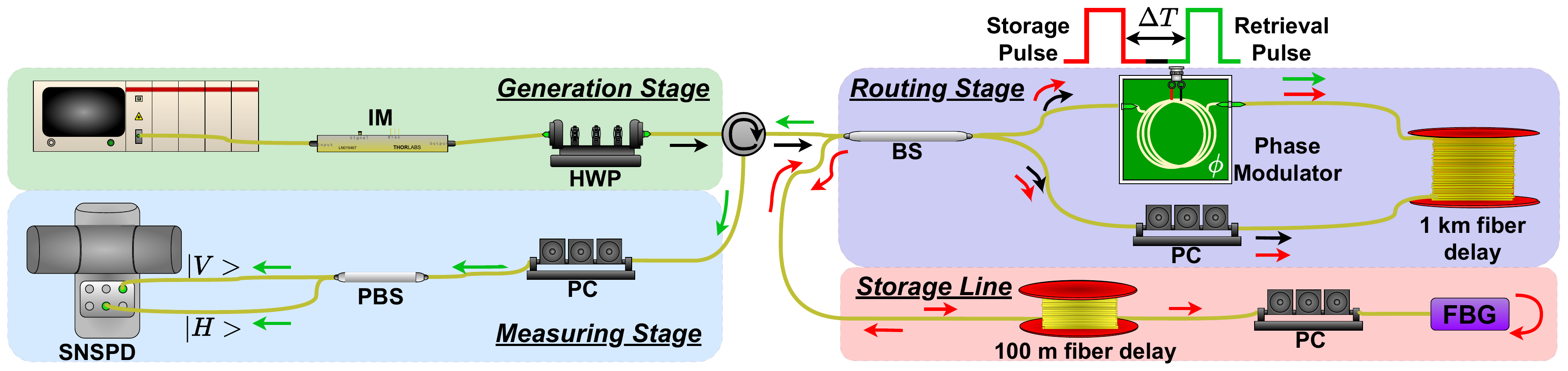}
    \caption{Experimental setup. Polarization quantum states produced at the generation stage are routed by the routing stage, which is driven the storage pulse applied through the poled fiber phase modulator, and then to the storage line. Then the routing stage retrieves back the states, following a retrieval pulse through the modulator, after an $\eta$ number of passes in the storage line. Please see text for details. Half-wave plate (HWP); Intensity modulator (IM), beamsplitter (BS), polarizing beamsplitter (PBS); fiber Bragg grating (FBG); polarization controller (PC); superconducting nanowire single photon detector (SNSPD). }
    \label{fig:expsetup}
\end{figure}

The experimental setup is presented in figure \ref{fig:expsetup}. A semiconductor laser diode at 1546.9 nm is connected to a fiber-pigtailed LiNbO$_3$ intensity modulator producing 50 ns long pulses at a repetition frequency of 1 kHz, and then to an attenuator (not shown for simplicity), generating weak coherent states at the single-photon level . A half-wave plate (HWP) is used to adjust the polarization of the quantum state between the orthogonal states $|H\rangle$, $|V\rangle$ and their linear superpositions. The quantum state passes through a circulator and into the routing stage, which is constructed with a poled phase modulator \cite{Camara15} and 1 km optical fiber delay (chosen for convenience).

\begin{figure}[H]
    \centering
    \includegraphics[trim = 0 39 0 50 , clip,width=0.994\linewidth]{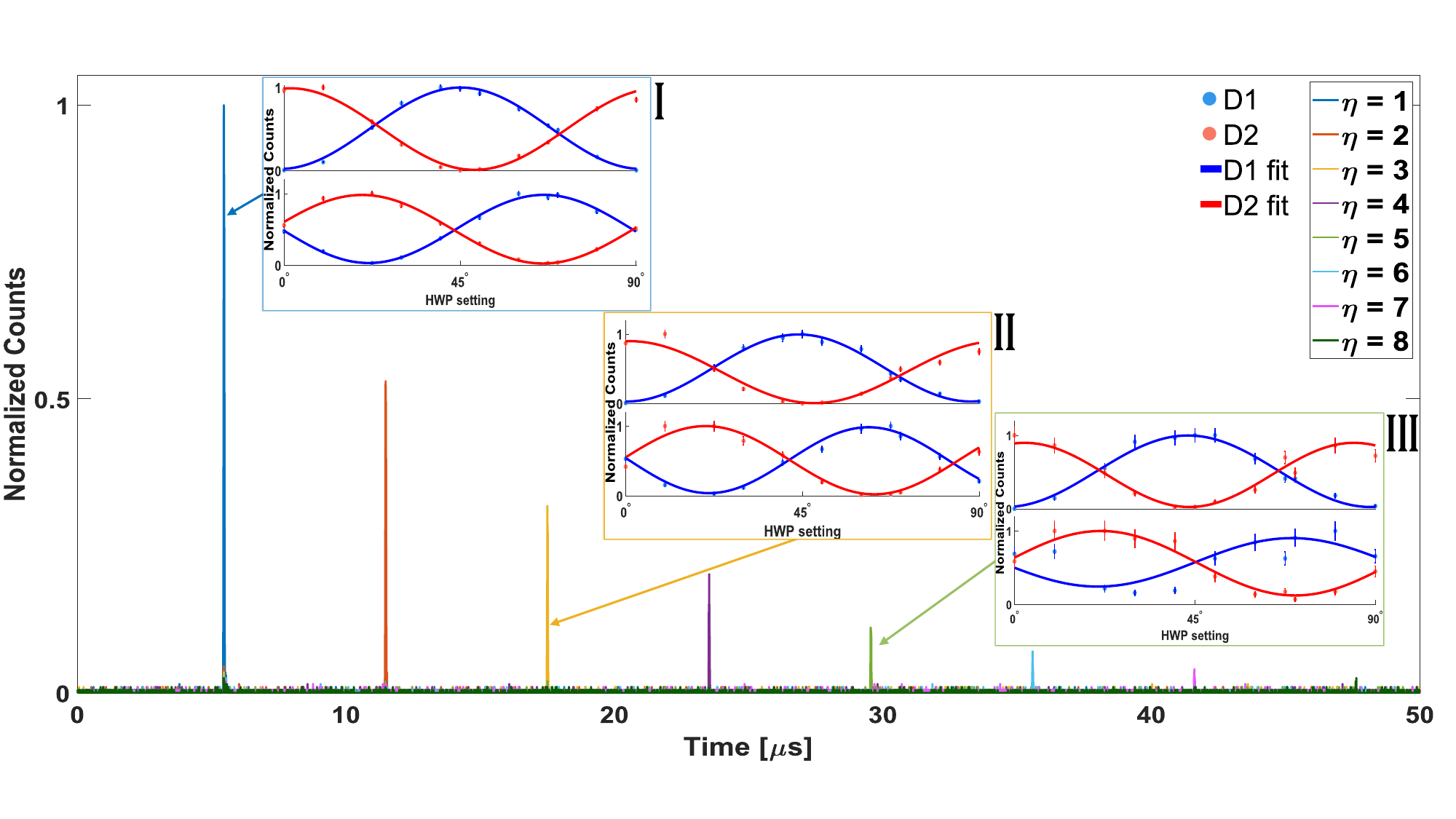}
    \caption{Experimental results showing the different retrieval times $\eta\Delta T$ of single-photons from the buffer. $\eta$ is the number of storage cycles in the memory loop and $\eta = 1$ corresponds to the photon directly retrieved without going to the storage line. Results from measuring polarization qubit storage is shown in the insets I, II and III, where normalized counts are shown.}
    \label{fig:results}
\end{figure}

\noindent A pulse generated by a function generator is sent to a high voltage driver (900 V and width of 180ns) to drive the poled fiber phase modulator (0.4 dB loss), which then routes the quantum state to the 100 m long storage line (red arrow in Fig. \ref{fig:expsetup}). It will then be reflected back by a fiber Bragg grating (FBG), and then back to the Sagnac loop which will also act as a mirror, reflecting the quantum state back to the storage line and so on. The 100 m fiber, delays the reflection to avoid unwanted memory read out. When a second high-voltage ``retrieval" pulse is applied to the poled fiber modulator, the quantum state will be routed out of the routing stage and back to the circulator and to the measuring stage (green arrow). The quantum state is projected by a fiber polarisation beamsplitter (PBS) onto its two orthogonal outputs, which are connected to two super conducting nanowire single photon detectors (SNSPDs, ID Quantique ID281, 90\% detection efficiency). A manual polarization controller (PC) before the PBS is used to adjust the measurement basis of the PBS between the computational ($|H\rangle, |V\rangle$) or the logical basis $[1/\sqrt{2}(|H\rangle \pm |V\rangle)]$. A first measurement, where the PBS is removed, is performed to detect the quantum state at different retrieval times $\eta\Delta T$, where $\eta$ is an integer and $\Delta T$ is the storage period of the buffer. The results, taken with a time tagger (IdQuantique ID1000) are shown in the main part of Fig. \ref{fig:results}, where we can observe up to 8 storage periods, corresponding to approx. 47 $\mu$s. In the second measurement (shown in the three insets in Fig. \ref{fig:results}) we measured the storage of polarization quantum states for the computational and logical bases (upper and lower graphs respectively) as a function of the HWP setting in the generation stage for three different retrieval times ($\eta = \{1, 3, 5\}$). We obtain average visibilities for both bases of $95.5 \%$, $95.3 \%$ and $83.5 \%$ respectively for the three different retrieval times, showing good storage preservation of the quantum states by the buffer.
\section{Conclusion and Discussion}
We have shown an all-fiber platform that can store and retrieve polarization quantum states at different time intervals. A crucial component is the low-loss poled fiber modulator, which enables the storage of the quantum states over several passes, a feat which would be much more limited with commercial telecom fiber-pigtailed fiber modulators ($\approx$ 3 dB). Our results opens possibilities for the storage of packets of several quantum states, which could make our system a key component of packet routing in future quantum networks. 

\bibliographystyle{unsrt} 
\bibliography{sample}

\end{document}